\documentclass[twocolumn, times, tighten]{aastex62}
\usepackage{graphicx}
\usepackage{contcaption,mathtools}
\usepackage{xcolor}

\journalinfo{Published in ApJS}

\shorttitle{WRGs from LoTSS DR2} \shortauthors{Bera et al.}

\begin{document}

\title{A morphological identification and study of radio galaxies from LoTSS DR2 -- I: the `Winged' Radio Galaxies}

\correspondingauthor{}

\author{Soumen Kumar Bera}
\affil{Department of Astronomy, Xiamen University, Xiamen, Fujian 361005, China}

\author{Taotao Fang}
\affil{Department of Astronomy, Xiamen University, Xiamen, Fujian 361005, China}
\email{fangt@xmu.edu.cn}

\author{Tapan K. Sasmal}
\affil{National Astronomical Observatories, Chinese Academy of Sciences, Beijing 100101, China}

\author{M. Kunert-Bajraszewska}
\affil{Institute of Astronomy, Faculty of Physics, Astronomy and Informatics, NCU, Grudziadzka 5, 87-100 Toru\'n, Poland}

\author{Xuelei Chen}
\affil{National Astronomical Observatories, Chinese Academy of Sciences, Beijing 100101, China}

\author{Soumen Mondal}
\affil{Department of Physics, Jadavpur University, Kolkata 700032, India}

\begin{abstract}
We conducted an extensive identification and analysis of various morphological classes and subclasses of radio galaxies using the latest high-resolution data from the second data release of the LOFAR Two-Metre Sky Survey (LoTSS DR2). This paper presents the first results of our large-scale investigation: a new catalog of ``winged" radio galaxies (WRGs). These objects represent a fascinating class of irregular radio galaxies, characterized by a pair of secondary radio lobes (``wings") in addition to the primary active lobes. We identified and cataloged 621 new WRGs and 403 additional candidates. Among the confirmed winged sources, 382 are classified as ``X"-shaped radio galaxies (XRGs), while the remaining 239 are ``Z"-shaped radio galaxies (ZRGs). We also estimated several basic parameters for these winged sources and performed a Fanaroff-Riley (FR) classification. Our results show that the majority of the sources ($\sim$88\%) exhibit edge-brightened radio lobes and high average radio power ($\rm log_{10}[P_{144MHz} / W Hz^{-1}]$ = 26.25), consistent with an FR-II classification. The average spectral index between 144 MHz and 1.4 GHz is --0.84, which is steeper than that found for previously identified winged sources based on higher-frequency data from the VLA Faint Images of the Radio Sky at Twenty-Centimeters (FIRST) survey. This indicates that our study is capable of detecting fainter sources. The median linear size of the winged sources, 498 kpc, confirms that these are large-scale structures, with approximately 16\% having sizes exceeding 0.7 Mpc, making them potential candidates for giant radio galaxies.
\end{abstract}


\section{Introduction}
\label{sec:intro}
Extragalactic radio sources are among the most powerful and energetic objects in the universe, with luminosities up to 100 times greater than those of star-forming galaxies and sizes extending to the megaparsec scale \citep{Ke88}. Active galactic nuclei (AGNs) are widely believed to host and reside at the centers of these radio sources \citep{Ko13}. Powered by supermassive black holes, AGNs release two oppositely directed relativistic jets of nonthermal radio emission, which extend from the central core as highly collimated outflows of ionized plasma. These jets transfer energy from the core to the outer lobes of the galaxy. Early literature primarily focused on classical double-lobed radio galaxies (e.g., \citet{Je53,Mo66,Wi78,Mi80}). However, subsequent observational studies revealed a wide variety of irregular morphologies, featuring complex structures and twisted formations that deviate from the classical configurations.

Irregular radio morphologies include several defined classes and sub-classes, as well as some unusual and miscellaneous structures. Among the defined classes are WRGs \citep{Le92}, bent-tail radio galaxies \citep{Ry68,Bl00,Sa22a}, and double-double radio galaxies \citep{La99,Sa06}. Another intriguing class is the recently identified odd radio circles (ORCs), introduced by \citet{No21}. Sub-classes of irregular radio morphologies include 'X'-shaped radio galaxies (XRGs) and 'Z'-shaped radio galaxies (ZRGs), both of which are discussed in detail later. Additionally, wide-angle tail galaxies, characterized by large opening angles in their radio jets \citep{Ei84}, and narrow-angle tail galaxies, distinguished by sharply curved jets \citep{Od86}, further expand the diversity of these structures. Beyond these classifications, irregular radio morphologies also encompass strange and miscellaneous structures \citep{Sa22b,Be24} and unspecified peculiar formations \citep{Va83, Bu96, Sh15}.

In addition to the classifications mentioned above, radio galaxies are commonly divided based on the historical morphological classification proposed by \citet{Fa74}. Less luminous galaxies, characterized by relatively diffuse and poorly collimated jets that terminate in diffuse lobes (known as edge-dimmed morphology), are classified as Fanaroff-Riley class I (FR-I) objects. In contrast, Fanaroff-Riley class II (FR-II) sources exhibit well-defined jets that maintain collimation over larger distances, often terminating in bright, sharply defined lobes (edge-brightened morphology). Interestingly, rare cases exist where double-lobed radio sources display distinct morphologies in each lobe, with one lobe showing FR-I characteristics and the other FR-II features. Such objects are termed HYbrid MOrphology Radio Sources (HYMORS; \citet{Go00,Ga06}). Finally, irrespective of these classifications, radio galaxies with linear size greater than 0.7 Mpc are categorized as giant radio galaxies (GRG; \citet{Is99,Ku18,Da20}).

The study of radio galaxy morphology is crucial for understanding various astrophysical phenomena. A fundamental connection between morphology and core accretion mechanisms has been proposed \citep{Hi79, La94, Mi17}. Morphology can also be influenced by environmental interactions, such as encounters with neighboring galaxies, interactions with the intracluster medium, or tidal forces within galaxy groups \citep{Wi02,Sa22a, Go23}. Radio galaxies are also supposed to have a signiﬁcant part in the metallization and magnetization \citep{Kr01, Go01, Ry08}. However, the mechanisms behind certain morphological classes, such as WRG \citep{Be22} and ORC \citep{No21}, remain unclear. Morphological studies are vital for advancing our understanding of galaxy evolution, AGN physics, galaxy formation, environmental effects, and cosmological implications. Given this background, we have done an extensive identiﬁcation and study on the different morphological classes and sub-classes of radio galaxies. In this paper (the first in a series) we present the results of our work on winged radio sources.

As mentioned earlier the ‘winged’ radio galaxy \citep{Le92, Ch07, Ya19, Be20, Be22} is a small exotic subset of a double-lobed radio galaxy that has an additional set of secondary lobes, aligned with the primary lobe pairs. The secondary lobes are found to be diffused and have low surface brightness. These secondary lobes are known as `wings' and hence the source `winged' radio source. Depending on the point of ejection of the wings from the primary lobes, WRGs are classified into two sub-classes, XRG \citep{Ch07, Ya19, Be20, Be22} and ZRG \citep{Zi05, Be20, Be22}. \citet{Ri72} presented the earliest known winged radio source, 3C 272.1, which exhibits a Z-shaped morphology. Subsequently, \citet{Le92} identified a total of 11 similar sources. \citet{Ch07} conducted the first systematic search for X-shaped radio galaxies at 1.4 GHz using the NRAO Very Large Array (VLA) Faint Images of the Radio Sky at Twenty-Centimeters (FIRST; \citet{Be95}), cataloging a total of 100 XRGs. With the latest data release of FIRST, \citet{Ya19} and \citet{Be20} identified 290 and 296 winged radio galaxy (WRG) sources, respectively. Additionally, \citet{Pr11}, using the same FIRST data, listed 156 XRG candidates (including 134 newly identified sources) through an automated morphological identification procedure. These identifications were primarily based on data at 1.4 GHz (i.e., $>$ 1 GHz). In the low-frequency domain, specifically below 300 MHz, only two studies have reported identifications. \citet{Bh22} discovered 58 winged sources using the first alternative data release of the TIFR Giant Metrewave Radio Telescope Sky Survey (TGSS ADR1; \citet{In17} at 150 MHz. Recognizing the limitations in resolution and sensitivity of TGSS ADR1, \citet{Be22} analyzed the LOFAR Two-Metre Sky Survey First Data Release (LoTSS DR1; \citet{Sh19}), identifying 26 new sources (40 in total). However, due to the limited sky coverage of LoTSS DR1 (424 deg$^2$), further studies using high-quality low-frequency data with broader sky coverage are essential.

WRGs have inspired various models to explain their origin. The most popular is the backflow of plasma model \citep{Le84, Ca02}, which suggests that wings form as plasma flows back toward the core from the hotspots. Alternatively, wings might result from lateral expansion of the surrounding medium along the minor axis, supported by observational evidence \citep{Ca02, Sa09}. Hydrodynamic simulations \citep{Ho11, Gi22} also validate this model. Another prominent model is the jet reorientation or spin-flip model \citep{Me02, Zi05}, proposing that jet reorientation, possibly due to supermassive black hole mergers, creates wings as relic emission along the previous axis. The formation of wings could be also explained by the presence of an unresolved pair of AGNs within the central core of the host galaxy. It is hypothesized that each AGN generates a pair of jets, which are ejected in different directions \citep{La05, La07}. While observational evidence \citep{La05, La07} supports this theory, it struggles to explain why primary lobes align with the optical major axis or why XRGs lack FR-II type lobes on both sides. Other theories include the buoyancy model \citep{Gu73, Ro01}, which posits that buoyant forces shape the wings but require specific interstellar medium or intergalactic medium conditions \citep{La07}, and the jet-shell interaction model \citep{Go83}. A unified framework \citep{Ga20} proposes that XRGs are transitional objects between retrograde and prograde black hole accretion states. Despite these models, systematic large-sample studies remain rare due to limited multi-frequency data. Most prior work focuses on individual case studies, leaving many questions about the properties and population-level attributes of WRGs unresolved.

To address the challenges and unanswered questions surrounding winged radio sources, the first step is to establish a bona fide, sufficiently large sample of such sources. While existing data from the mid-to-high frequency FIRST survey at 1.4 GHz provides a foundation, identifying additional sources from low-frequency data is crucial. Once a comprehensive dataset spanning both low and high frequencies is compiled, deeper studies of these sources can be conducted in future work.

In this paper, we present an extensive search for winged radio sources using the LOFAR Two-Metre Sky Survey Second Data Release (LoTSS DR2; \citet{Sh22}) at 144 MHz. The paper is structured as follows: Section \ref{sec:data} discusses the LoTSS DR2 survey data and the rationale for its selection. Section \ref{sec:method} outlines the source identification methodology. Section \ref{sec:result} presents the search results, along with a brief analysis of the identified sources and their properties. Finally, Section \ref{sec:summary} provides a summary and an overview of future work in this whole morphological study.

In our work, we have used the following cosmology parameters: $\rm H_0 = 67.4$ km s$^{-1}$ Mpc$^{-1}$, $\rm \Omega_m = 0.315$, and $\rm \Omega_{vac} = 0.685$ \citep{Pa20}. The definition, we used for the spectral index ($\alpha$) is $\rm F_{\nu} \propto \nu^{\alpha}$, where $\rm F_{\nu}$ is the particular flux density at the frequency $\nu$.

\section{The Survey Data: LoTSS DR2}
\label{sec:data}
The Low-Frequency Array (LOFAR; \citet{Va13}) is a state-of-the-art radio telescope system designed to explore the universe at low radio frequencies, ranging from 10 to 240 MHz. Its groundbreaking capabilities are exemplified by the ongoing LOFAR Two-Metre Sky Survey (LoTSS), which operates in the 120--168 MHz frequency range using the LOFAR High Band Antenna (HBA) \citep{Sh17}. LoTSS aims to map the entire northern sky at LOFAR’s full 6-arcsecond resolution with a sensitivity of approximately 100 $\mu$Jy beam$^{-1}$, although sensitivity varies with declination. At its central frequency of 144 MHz, LoTSS delivers sensitivity $\sim$1.5 times greater than the FIRST survey (angular resolution of 5$^{\prime\prime}$ at 1.4 GHz), previously the highest-resolution sky survey. This sensitivity and resolution enable LoTSS to identify around 10 million radio sources, predominantly star-forming galaxies and active galactic nuclei (AGNs), ensuring it remains a vital resource well into the Square Kilometer Array (SKA) era.

\begin{figure*}
\hspace*{0.75cm}
\vbox{
\centerline{
\includegraphics[angle=0,height=7.9cm,width=7.9cm]{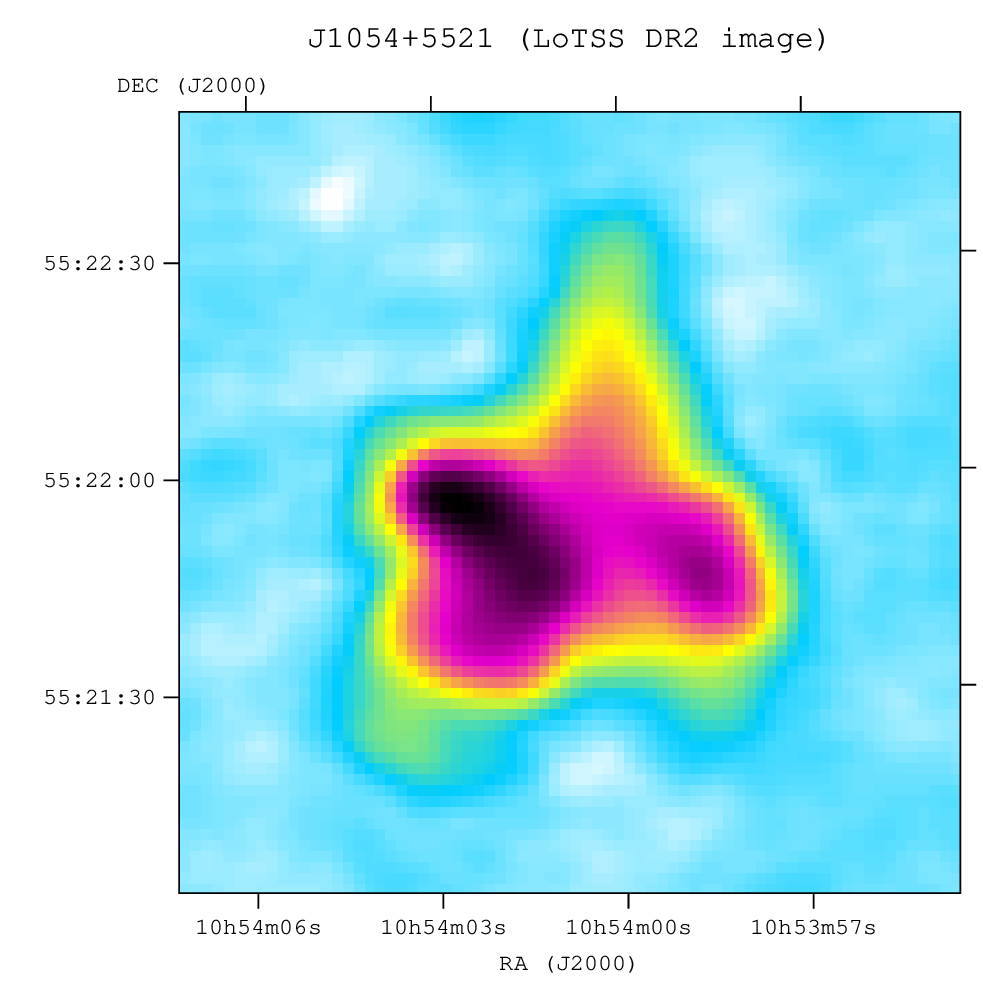}
\includegraphics[angle=0,height=7.9cm,width=7.9cm]{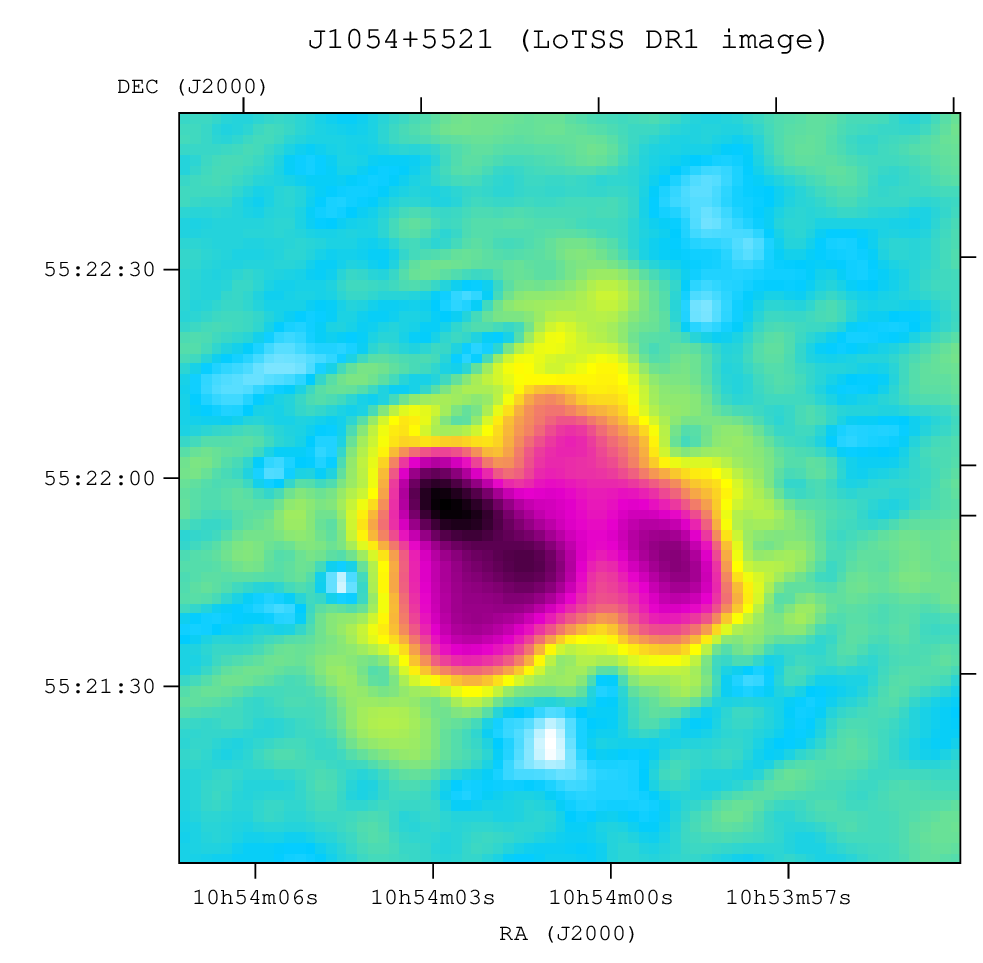}}}
\vskip 2.5cm
\caption{The images illustrate the improvement in data quality between the LoTSS DR2 (left panel) and the LoTSS DR1 (right panel).}
\label{fig:compr}
\end{figure*}

The second data release (DR2) of LoTSS (LoTSS DR2) marks a significant milestone, further enhancing the survey's capabilities. Building upon the success of its predecessor, DR1, DR2 expands the survey area to approximately 5,700 deg$^2$, covering thirteen times the area of DR1’s 424 deg$^2$. This expanded coverage includes the DR1 region, which had 63 pointings, as part of a total 841 pointings distributed across two adjacent regions (Table \ref{table:specs}). The first region, centered at 12h45m +44$^\circ$30$^\prime$, spans 4,178 deg$^2$ with 626 pointings, while the second, centered at 1h00m +28$^\circ$00$^\prime$, covers 1,457 deg$^2$ with 215 pointings. Beyond its broader coverage, LoTSS DR2 introduces significant improvements in data quality. The data reduction process has been refined and optimized, with enhanced direction-dependent calibration routines and improved image processing techniques. The flux density scale has been meticulously refined during the mosaicing process, resulting in greater accuracy. Major advancements in DR2 include improved dynamic range and enhanced fidelity of faint diffuse emissions, enabling the detection of subtle astrophysical features \citep{Ta21, Sh22}. These advancements are well illustrated by a comparison of images of a winged source J1054+5521 from DR1 and DR2 (see Figure \ref{fig:compr}). The DR2 image is smoother, with more prominent and better-resolved lobes. The angular size and flux density of the source also show marked improvement, increasing from 60 arcsec and 1,604.0 mJy in DR1 to 74 arcsec and 1,798.8 mJy in DR2, respectively. Thus, LOFAR's unique design, particularly its short baselines, is critical for imaging large-scale isotropic morphological structures such as extended radio lobes, making it the optimal database for the current study.

\begin{table}
\begin{center}
\caption{A short specification of LoTSS DR2}
\begin{tabular}{ll}
\hline\hline
Total sky coverage:	& 5634 deg$^2$; 27\% of the northern sky \\
Number of pointings:	& 841\\
Survey regions: 	& (i) 12h45m +44$^\circ$30$^\prime$; \\
(two areas centered at) & area: 4178 deg$^2$; pointings: 626 \\
	                & (ii) 01h00m +28$^\circ$00$^\prime$; \\
			& area: 1457 deg$^2$; pointings: 215 \\
Survey timeline:	& 3451 h  \\
Integration time: 	& $\sim$16 h \\
Frequency range:        & 120 -- 168 MHz \\
Central frequency:	& 144 MHz \\
Resolution:		& 6$^{\prime\prime}$ \\
Median rms sensitivity:	& 83 $\mu$Jy beam$^{-1}$ \\
Number of sources:	& 4,396,228 \\
\hline
\end{tabular}
\label{table:specs}
\medskip\\
\end{center}
\end{table}

\begin{table*}
\caption{ The winged radio sources from LoTSS DR2}
\label{table:samplet}
\begin{centering}
\scriptsize
\begin{tabular}{ccllccccccccccccc}
\hline\hline
Cat & Short  &~~~~~R.A.    &~~~~~Decl.    & $z$& $z_{type}$&$m_{r}$ &$M_{R}$&$F_{144}$&Isl\_rms&$F_{1400}$&$\alpha_{144}^{1400}$&$\theta$&$l$  &$P_{144 MHz}$   &XRG/&FR-     \\
No. & Name   &~~(J2000.0)  &~~(J2000.0)   &      &          &        &       &  (mJy)  &(mJy/   & (mJy)    &                     &(arcsec)&(kpc)&($\times10^{26}$&ZRG &type   \\
    &        &             &              &      &          &        &       &         & beam)  &          &                     &        &     & W/Hz)          &    &(I/II) \\
\hline
~~1 & J0001+2821 & 00 01 53.01    & +28 21 38.4            &  0.41   & PHOT  & 19.6 & --22.8  & ~562.6  & 0.158 & ~48  & --1.08 & ~99  & ~551 &  3.60   & XRG   & ---   \\ 
~~2 & J0003+2258 & 00 03 02.4     & +22 58 39$^{\ast}$     &  ---    & ---   & ---  &  ---    & ~~56.1  & 0.094 & ~14  & --0.61 & ~82  &  --- &  ---    & ZRG   & ---   \\
~~3 & J0003+1810 & 00 03 07.31    & +18 10 01.6            &  ---    & ---   & 21.4 &  ---    & 3354.7  & 1.076 & 449  & --0.88 & 122  &  --- &  ---    & XRG   & II    \\ 
~~4 & J0004+3702 & 00 04 05.6     & +37 02 21$^{\ast}$     &  ---    & ---   & ---  &  ---    & ~446.1  & 0.175 & ~78  & --0.77 & ~61  &  --- &  ---    & XRG   & II    \\ 
~~5 & J0008+3502 & 00 08 11.29    & +35 02 18.3            & ~0.145  & SPEC  & 17.2 &  ---    & ~583.3  & 0.334 & 119  & --0.70 & ~53  & ~140 &  0.34   & XRG   & II    \\ 
... & .....      & .....          & .....                  & .....   &.....  &..... & .....   &.....    &.....  &..... & .....  &..... &..... &.....    &.....  &.....  \\
... & .....      & .....          & .....                  & .....   &.....  &..... & .....   &.....    &.....  &..... & .....  &..... &..... &.....    &.....  &.....  \\
... & .....      & .....          & .....                  & .....   &.....  &..... & .....   &.....    &.....  &..... & .....  &..... &..... &.....    &.....  &.....  \\
617 & J2346+3053 & 23 46 12.22    & +30 53 31.1            &  0.43   & PHOT  & 18.8 & --24.0  & ~208.2  & 0.104 & ~44  & --0.68 & ~57  & ~328 &  1.31   & XRG   & II    \\ 
618 & J2348+3142 & 23 48 40.85    & +31 42 33.0            & ~0.469  & SPEC  & 19.9 & --23.1  & ~239.1  & 0.096 & ~43  & --0.75 & 124  & ~757 &  1.92   & XRG   & II    \\ 
619 & J2350+2747 & 23 50 53.42    & +27 47 08.8            &  0.69   & PHOT  & 21.3 & --23.1  & ~375.8  & 0.165 & ~28  & --1.14 & ~69  & ~509 &  9.15   & XRG   & ---   \\ 
620 & J2351+2841 & 23 51 56.94    & +28 41 56.2            &  1.59   & PHOT  & 24.6 &  ---    & ~~47.0  & 0.122 & ~~8  & --0.78 & ~47  & ~409 &  6.61   & ZRG   & I     \\ 
621 & J2359+1706 & 23 59 11.1     & +17 06 11$^{\ast}$     &   ---   &  ---  &  --- &   ---   & 3601.4  & 0.167 & 431  & --0.93 & 584  &  --- &  ---    & XRG   & ---   \\ 
\hline
\end{tabular}
\end{centering}
\scriptsize
Notes:\\
Notes: This is a sample excerpt from the table, which will be available in its entirety in the online link of the published version.\\
Column information of the Table: Column 1 is the catalog number, Column 2 is the short name (IAU source identifier as JHHMM+DDMM), Columns 3 and 4 have the Right Ascension (R.A. in hh mm ss.ss), and Declination (Decl. in dd mm ss.s), both in J2000.0. The optical host position of the sources is presented here. If no optical host is found for a particular source, then we use radio coordinates for the source. The radio positions are given with less precision (R.A. as hh mm ss.s and Decl. as dd mm ss). Columns 5 and 6 represent the redshift ($z$) and the redshift type ($z_{type}$), i.e.; either spectroscopic (SPEC) or photometric (PHOT). The spectroscopic redshift values are taken from the Sloan Digital Sky Survey Data Release 16 (SDSS DR16: \citet{Ah20}) and the Dark Energy Spectroscopic Instrument (DESI: \citet{Le13}). The photometric redshifts are taken from \citet{Du22}. The PHOT redshifts are given with less precision, up to 0.01 place. The r-band magnitude ($m_{r}$) and the rest frame magnitude (r-band; $M_{R}$) are given in Columns 7 and 8, respectively. The next column is the integrated radio flux of the source at frequencies 144 MHz ($F_{144}$ in mJy). The respective rms noise in the island (Isl\_rms) for each of the sources is given in Column 10. The estimated NVSS flux at 1400 MHz ($F_{1400}$ in mJy) is listed in Column 11. Here, we use the LoTSS DR2 flux for $F_{144}$, and for $F_{1400}$ we use the NVSS flux, due to its better flux detection ability. The $F_{144}$ values are taken from \citet{Ha23}, and in some cases, we measure the flux value manually if the particular flux is not available in the catalog.  We present the two-point spectral index $\alpha_{144}^{1400}$ in Column 12. The angular size ($\theta$) and the linear size ($l$) of the sources (in kpc) are given in Column 13 and Column 14, respectively. The radio power ($P_{144 MHz}$ in W Hz$^{-1}$) is listed in Column 15. The winged sources' classification--XRG or ZRG--is shown in the following column. The last column has the FR-classification.\\
$^{\dag}$ The source is present in \citet{Pr11}.\\
The coordinates with $^{\ast}$ marks indicate that they are radio coordinates; otherwise, they are the respective optical host position.\\
The $^{\star}$ mark on the spectroscopic redshift implies that the redshift is taken from DESI; otherwise, it is taken from SDSS.
\end{table*}

\section{The Methodology}
\label{sec:method}
\subsection{Identification procedure}
LoTSS DR2, which also encompasses the DR1 area, contains a total of 4,396,228 radio sources, marking the largest dataset from any radio survey to date \citep{Sh22}. In this study, we did not exclude the DR1 region from our search, as the improvements in data quality between DR1 and DR2 make it worthwhile to re-examine previously identified winged sources. Re-checking these sources not only validates our current methodology but also ensures that our analysis encompasses the full extent of LoTSS DR2 data available to date. \citet{Be22} previously identified 40 winged sources in the DR1 data, including 26 newly discovered ones. All of these 40 sources have also been identified in DR2 through the procedure described in this study and are presented in a separate table for reference and comparison, underscoring the consistency of our methodology and the enhanced reliability of the DR2 data. This comprehensive approach ensures a robust analysis of winged sources while building upon prior research.

Given the enormous size of LoTSS DR2 and the presence of numerous multi-component sources, it is neither feasible nor necessary to inspect every source visually. Instead, we employed a filtering strategy to identify sufficiently large sources exhibiting fundamental morphological characteristics. Following a method similar to \citet{Be22}, we established a size threshold for candidate sources. The smallest winged source identified in DR1 had an angular size of approximately 40$^{\prime\prime}$, so we set our lower size limit to half of this value, i.e., 20$^{\prime\prime}$. This threshold ensures that the minimum angular size of our selected sources is at least three times the convolution beam size (6$^{\prime\prime}$). The major axis of the fitted Gaussian was used as the measure of source size.

To maximize the comprehensiveness of our sample, we imposed no additional constraints on peak or integrated flux, nor on the dynamic range of the sources. This approach resulted in a sample of 204,789 sources meeting the size criterion. Each of these sources was visually examined to identify potential winged structures.

The identification and confirmation of winged sources were based on their contour maps. Contours were drawn starting from at least three times the rms noise level in the island (Isl\_rms; \citet{Ha23}), with the lowest contour typically above 2.5 $\times$ 10$^{-4}$ Jy. This ensures the reliability of the resulting radio structures. The corresponding .ﬁts images for each source were obtained from the LoTSS DR2 cutout service\footnote{https://lofar-surveys.org/dr2\_release.html}.

\subsection{Differentiating the winged radio sources and the candidates for winged radio sources}
Winged radio sources documented in studies using FIRST (e.g., \citet{Ch07, Ya19, Be20}) and LoTSS DR1 \citep{Be22} reveal that not all such sources exhibit secondary lobes on both sides. Furthermore, some winged candidates lack clearly discernible wings at specific frequencies. Various factors can obscure the visibility of wing structures, including the source's position and orientation relative to the viewing plane, as well as the directional offset of the wings from the primary lobe pair. Projection effects can exacerbate these issues, as wings may appear less distinct due to artifacts associated with older radio counterparts \citep{Ya19}. Observational limitations, such as insufficient resolution and sensitivity, also contribute to challenges in detecting wings, which often exhibit diffuse, low-surface-brightness emissions \citep{Ch07}. For instance, sources with marginal wing evidence in \citet{Ch07} display prominent wings in the higher-quality data of \citet{Ro18}. Additionally, Doppler boosting may unevenly amplify the flux of one side of a lobe pair, diminishing the visibility of the opposing lobe.

To ensure a comprehensive compilation, we account for both confirmed winged sources and plausible candidates. In this study, we created two distinct lists: one for bona fide winged radio sources and another for candidate winged sources. A source is classified as a winged radio source if it has prominent secondary lobes on both sides of the primary lobes. Here, ``prominent" secondary lobes are defined as those distinguishable in both radio images and contour maps and possessing sufficient size relative to the primary lobes. While the secondary lobes do not need to be symmetric, they must be significantly visible on both sides.

This primary list comprises confirmed winged sources. The remaining sources, which do not meet these strict criteria, are included in the candidate list. Candidates encompass sources with one-sided wings, those with insufficient wing size, or wings that are not clearly recognizable in radio images or contour maps. This classification ensures a thorough and methodical approach to identifying winged sources while accounting for observational and intrinsic limitations.

\begin{figure*}
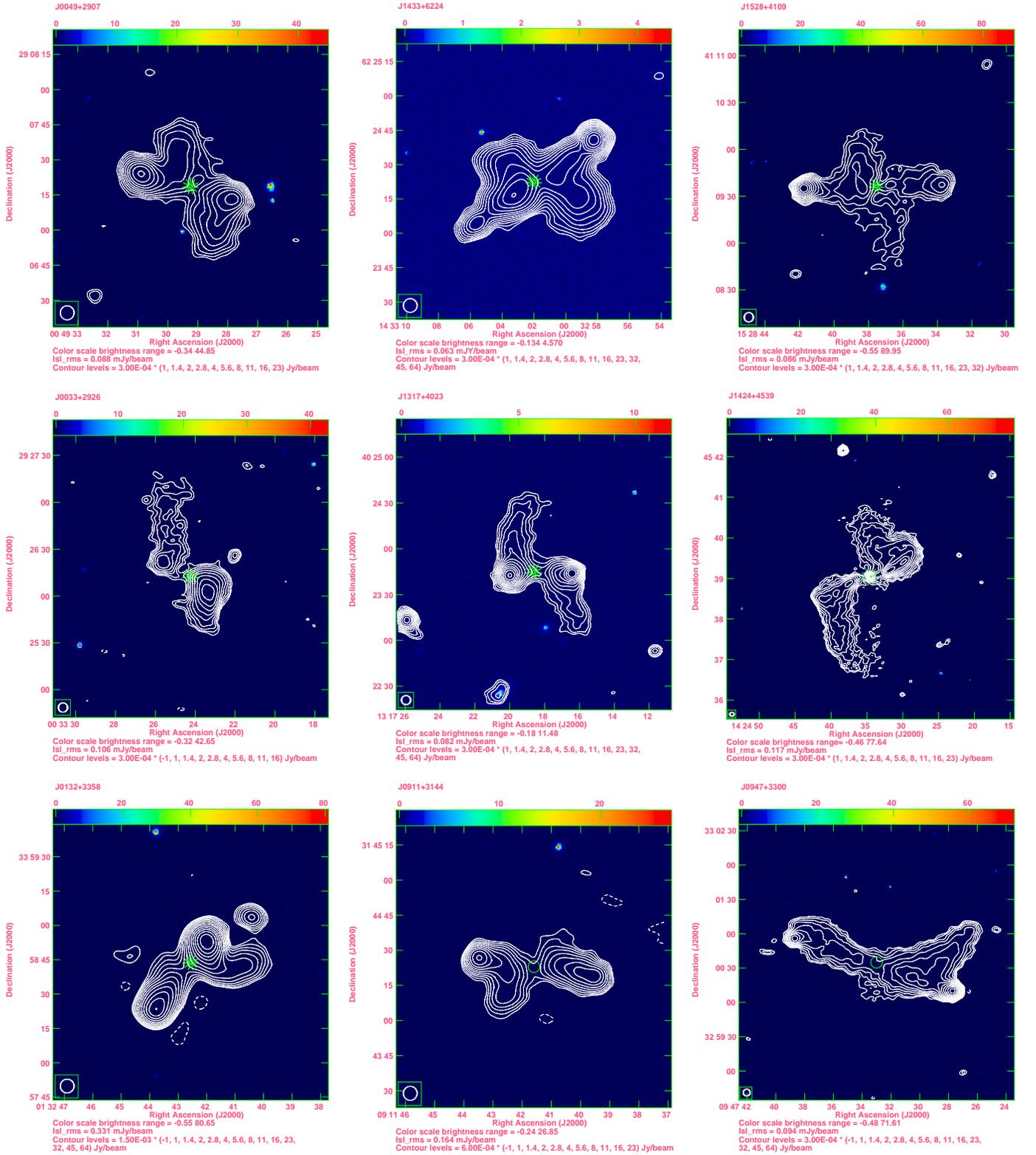

\vbox{
\centerline{
\includegraphics[angle=0,height=7.0cm,width=6.1cm]{./J0049+2907X.PS}
\includegraphics[angle=0,height=7.0cm,width=6.1cm]{./J1433+6224X.PS}
\includegraphics[angle=0,height=7.0cm,width=6.1cm]{./J1528+4109X.PS}}}
\vbox{
\centerline{
\includegraphics[angle=0,height=7.0cm,width=6.1cm]{./J0033+2926Z.PS}
\includegraphics[angle=0,height=7.0cm,width=6.1cm]{./J1317+4023Z.PS}
\includegraphics[angle=0,height=7.0cm,width=6.1cm]{./J1424+4539Z.PS}}}
\vbox{
\centerline{
\includegraphics[angle=0,height=7.0cm,width=6.1cm]{./J0132+3358P.PS}
\includegraphics[angle=0,height=7.0cm,width=6.1cm]{./J0911+3144P.PS}
\includegraphics[angle=0,height=7.0cm,width=6.1cm]{./J0947+3300P.PS}}}
\caption{Example sets of three XRG (top), ZRG (middle), and WRG candidates (bottom) from our identification. The LOFAR full-resolution 6$^{\prime\prime}$ radio images are displayed. Radio contours, plotted in $\sqrt2$ intervals, are overlaid on SDSS r-band images shown in color scale. The contours start at three times (or more) their respective Isl\_{rms}, increasing by a factor of 2. A star marks the optical host position, while a circle indicates the radio position when optical information is unavailable.}
\label{fig:sampleimg}
\end{figure*}

\subsection{Separating the XRGs and ZRGs}
The XRGs and the ZRGs are considered sub-classes of winged sources, distinguished by the ejection points of their secondary lobes relative to the primary lobes. In XRGs, the wings appear to emanate from the central region, giving the source an overall ``X" shape. Conversely, in ZRGs, the wings emerge from the outer edges of the primary lobes, resulting in a ``Z" shape.

For clarity, if a ridge line is drawn between the two secondary lobes, this line would pass through the central active galactic nucleus (AGN) in the case of XRGs. In ZRGs, however, the lateral offset of the wings means the ridge line typically does not intersect the central AGN.

In our classification, winged sources are identified as XRGs when their secondary lobes are ejected within the inner quarter of the primary jet's length, measured from the central core. In contrast, sources where the wings originate from the edges or beyond this one-fourth region are categorized as ZRGs. This classification applies exclusively to bona fide winged sources to ensure accuracy and consistency in identifying these sub-classes.

\section{Results and Discussions}
\label{sec:result}
Our comprehensive and detailed search for winged sources using LoTSS DR2 has identified a total of 621 new winged radio sources and 403 candidates. A sample of the winged sources is provided in Table \ref{table:samplet} and candidates are presented in Table \ref{table:candidate}. These tables are organized in ascending order of right ascension (RA) and are available in their entirety online only. The catalog excludes the 40 winged sources previously identified from LoTSS DR1 by \citet{Be22}. Additionally, we excluded any sources already identified in FIRST-based studies by \citet{Ch07}, \citet{Ya19}, and \citet{Be20}, including those overlapping with LoTSS DR2. These exclusions amount to a total of 78 sources, which are listed separately in Table \ref{table:common} in the Appendix. However, we retained the FIRST-based sources cataloged by \citet{Pr11}, as their morphologies have not yet been visually confirmed using radio maps.

Of the 621 winged sources identified, 382 exhibit X-symmetry (XRGs), while 239 display Z-symmetry (ZRGs), resulting in an XRG-to-ZRG ratio of 1.6. For comparison, \citet{Sa18} reported a ratio of approximately 2 for the sample from \citet{Ch07}. The higher number of ZRGs in our study suggests that our analysis is more sensitive to fainter sources, as ZRGs tend to be less luminous than XRGs.

No classification has been made for the 403 candidate sources. This group predominantly includes sources with one-sided wings or those with small or marginal evidence of wings. Approximately 40\% of the candidates exhibit one-sided wings, while nearly half feature small wings. Some sources in this group show small wings on both sides.

Figure \ref{fig:sampleimg} provides examples of three X-shaped radio galaxies (XRGs), three Z-shaped radio galaxies (ZRGs), and three candidates for winged sources. The radio images (from LoTSS DR2) are shown as contours overlaid on their respective SDSS (Sloan Digital Sky Survey; \citet{Gu06}) grayscale images.

Combining the winged sources and candidates, we identified a total of 1,024 new sources in the 5,700 deg$^2$ sky area covered by LoTSS DR2, resulting in an observed source density of $\sim$0.18 sources per square degree. This count significantly surpasses the identification efficiency from FIRST-based studies, such as \citet{Ch07}, \citet{Ya19}, and \citet{Be20}, which yield an observed source density of $\sim$0.06 sources per square degree when combined.

Hence, our identification rate is approximately three times higher than all previous FIRST-based studies combined. Notably, if we had included the previously identified sources, the source count would have been even higher. This better efficiency of LoTSS in finding winged sources is due to the fact that it is about 1.5 times more sensitive than FIRST, which makes the radio emission with a steep spectral index (alpha $\sim$--1) have a sufficiently high signal-to-noise ratio ($\sim$7) to be clearly detected.

The optical identification and redshift information for our sources were obtained from \citep{Ha23}, where 85\% of the sources have either optical or infrared identifications, with 58\% having redshift data. This translates to redshift information for approximately 72\% (449 sources) of our winged sources. Among these, 291 redshifts are photometric (PHOT), and 158 are spectroscopic (SPEC). Of the spectroscopic redshifts, 150 are sourced from the SDSS Data Release 16 (SDSS DR16; \citet{Ah20}), and the remaining 6 are from the Dark Energy Spectroscopic Instrument (DESI; \citet{Le13}). The redshifts range from 0.066 to 1.64, with a median value of 0.519$\pm$0.296.

For the 403 candidate winged sources, 174 have associated redshift data, including 120 photometric and 54 spectroscopic redshifts. Of the spectroscopic redshifts, 53 are from SDSS and one from DESI. The redshifts for candidates span a range of 0.00016 to 1.22, with a median value of 0.499$\pm$0.284.

In comparison, the median redshifts for previously identified winged sources from FIRST are $\sim$0.25, 0.37, and 0.25 as reported by \citet{Ch07}, \citet{Ya19} (for strong candidates), and \citet{Be20}, respectively. Comparing the redshift values and distributions, it is evident that the sources in both of our lists are more distant and have a more orderly distribution than those identified in earlier studies. This difference can be attributed to the superior sensitivity and resolution of the LoTSS DR2 data, as well as our more comprehensive and inclusive source selection criteria, which allow for the identification of fainter and more distant sources.

We also performed a basic analysis of our sources based on their r-band apparent magnitude ($\rm m_{r}$) and rest-frame r-band absolute magnitude ($\rm M_{R}$), where available. For the newly identified winged sources, the mean and median $\rm m_{r}$ values are approximately 20.2$\pm$2.0, while the corresponding $\rm M_{R}$ values are --23.0$\pm$0.7. In comparison, the mean and median $\rm m_{r}$ values for winged sources identified from FIRST observations \citep{Ch07, Ya19, Be20} are 19.0$\pm$2.2 and 19.3$\pm$2.2, respectively. This indicates that the r-band brightness of the new winged sources presented in this paper is significantly weaker than that of previously identified sources.

\begin{figure}
\includegraphics[angle=-90,width=9.0cm]{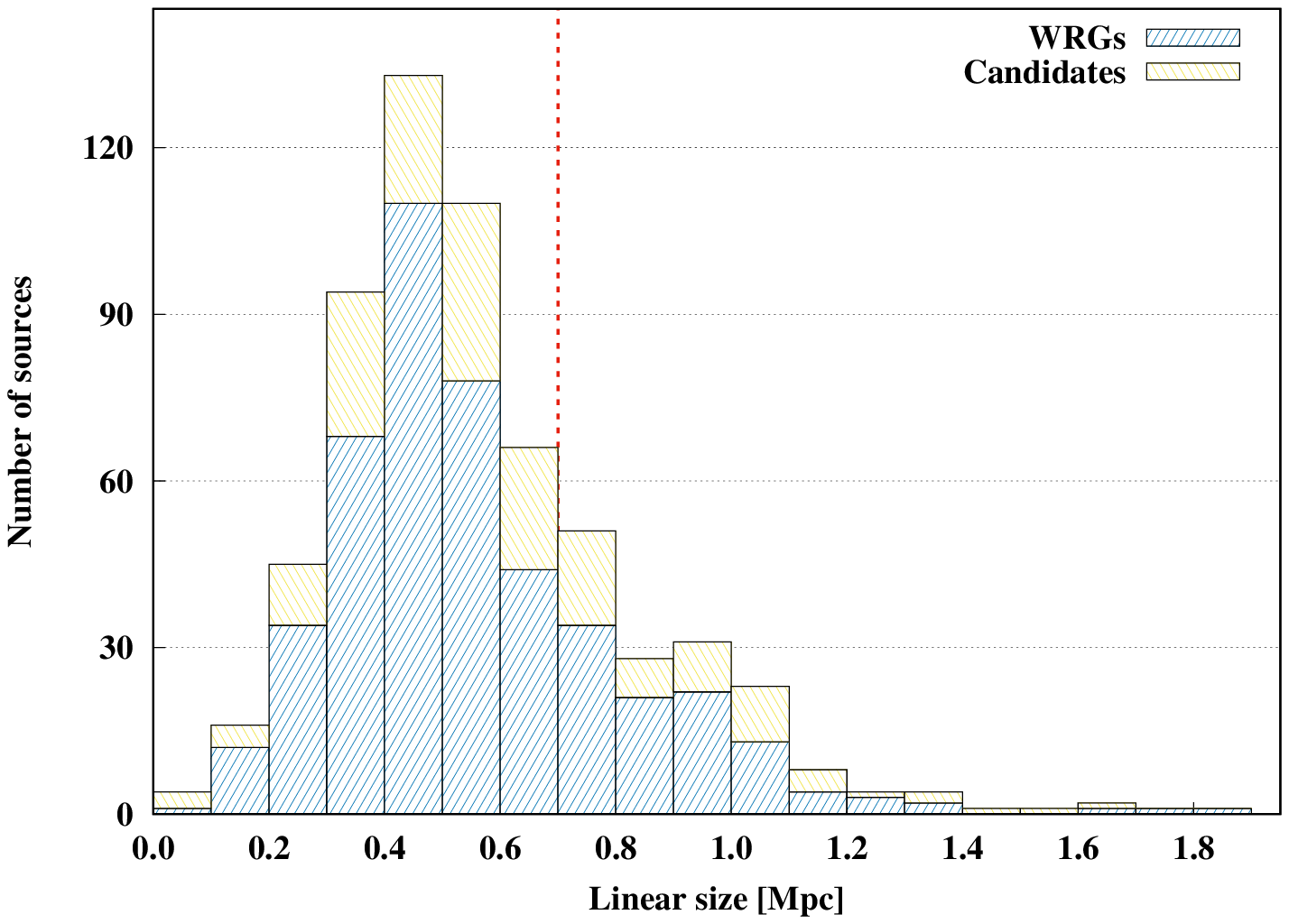}
\caption{Distribution of linear sizes among the identified winged sources and candidates. The vertical red dashed line at 0.7 Mpc marks the threshold for GRG candidates, which are located to the right of this line. The bin size is set to 0.1 Mpc.}
\label{fig:lsize}
\end{figure}

Among the population of winged sources, the median projected linear size (calculated for sources with available measurements) is 498$\pm$248 kpc. The maximum angular size is determined as the angular extent of the source, measured using the AIPS task ‘tvdist’. The source size is defined based on the lowest contour level. A total of 102 sources have a linear size exceeding 0.7 Mpc, making nearly 16\% of the total population candidates for giant radio galaxies (GRGs). Additionally, 25 sources have linear sizes greater than 1 Mpc, with the largest winged candidate, J1155+4029, reaching a linear size of approximately 1.8 Mpc. For the candidate winged sources, the median linear size is 558$\pm$288 kpc, with 53 sources (13\%) qualifying as potential GRG candidates. A histogram showing the distribution of linear size for the winged sources and the candidates is presented in Figure \ref{fig:lsize}. Comparing to previous studies, there are four XRGs with giant sizes in the \citet{Ya19} list, while \citet{Be20} and \citet{Be22} identified 34 and 13 giant XRGs, respectively. These findings indicate that the presence of wings in giant radio galaxies is not a rare phenomenon. However, the proportion of giant-sized winged sources is notably higher in LoTSS compared to FIRST, reflecting the superior sensitivity of the LoTSS survey. Among the winged sources with giant sizes, the median linear size is 0.92$\pm$0.20 Mpc. The presence of a subset of winged sources with giant dimensions suggests that some winged sources may contribute to the population of giant radio galaxies. This observation raises the possibility of an underlying triggering mechanism that could simultaneously drive the formation of wings and lead to the large linear sizes observed in giant sources.

In the following subsections, we present and discuss various physical parameters of the newly identified sources, focusing exclusively on the 621 winged sources, further divided into XRG and ZRG objects. Additionally, we compare the radio properties of winged sources identified in FIRST observations at 1.4 GHz with those found at the lower frequency of 144 MHz in the LoTSS survey. The FIRST sample includes all sources identified by \citet{Ch07}, \citet{Ya19}, and \citet{Be20}, while the LoTSS sample comprises sources from LoTSS DR1 \citep{Be22} alongside the new sources presented in this paper. Throughout the following sections, these groups will be referred to as the FIRST sample and the LoTSS sample, respectively. It is important to note that the number of sources included in specific analyses may vary depending on the availability of the required data for particular considerations and calculations.

\begin{figure*}
\includegraphics[angle=-90,width=9.0cm]{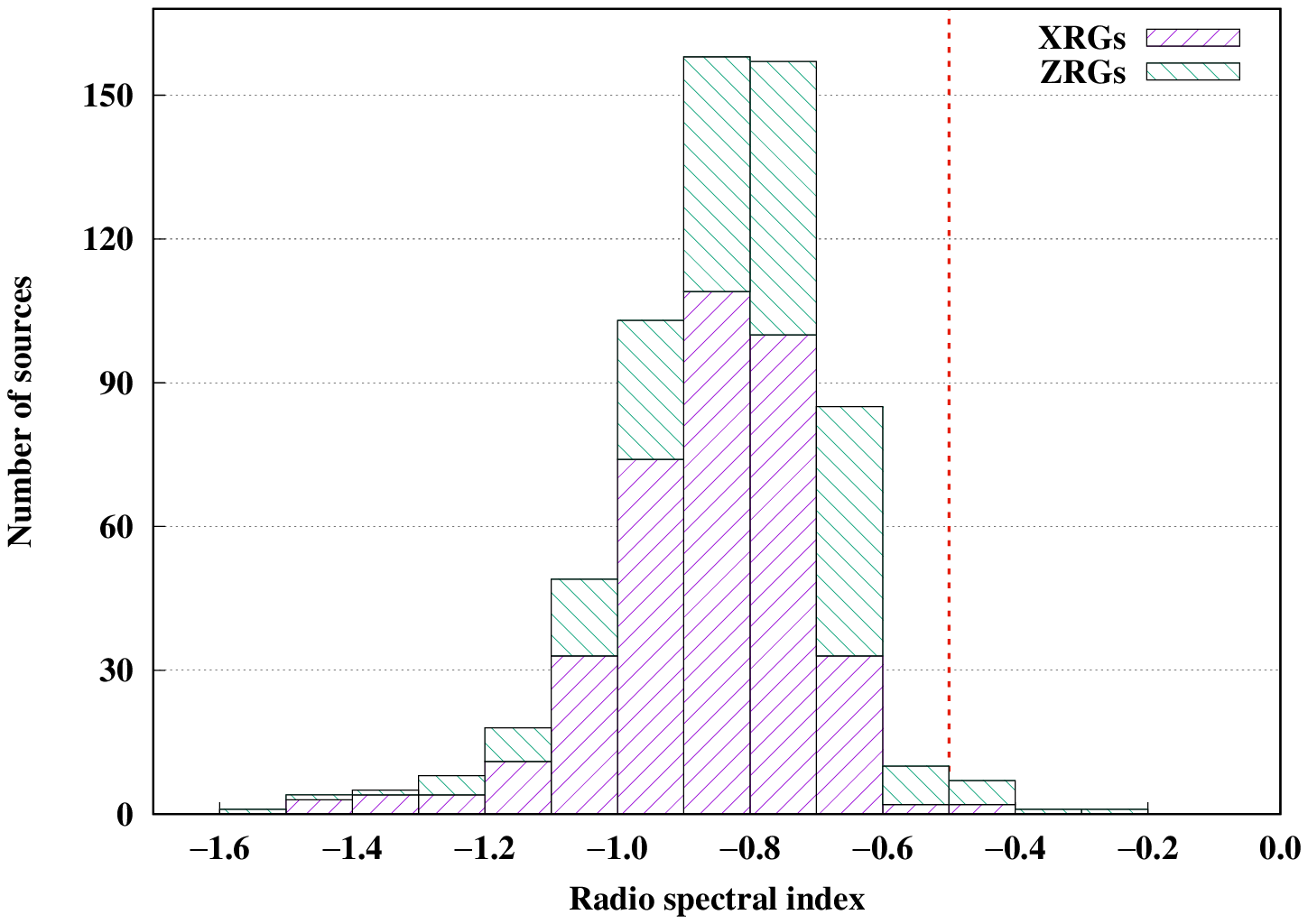}
\includegraphics[angle=-90,width=9.0cm]{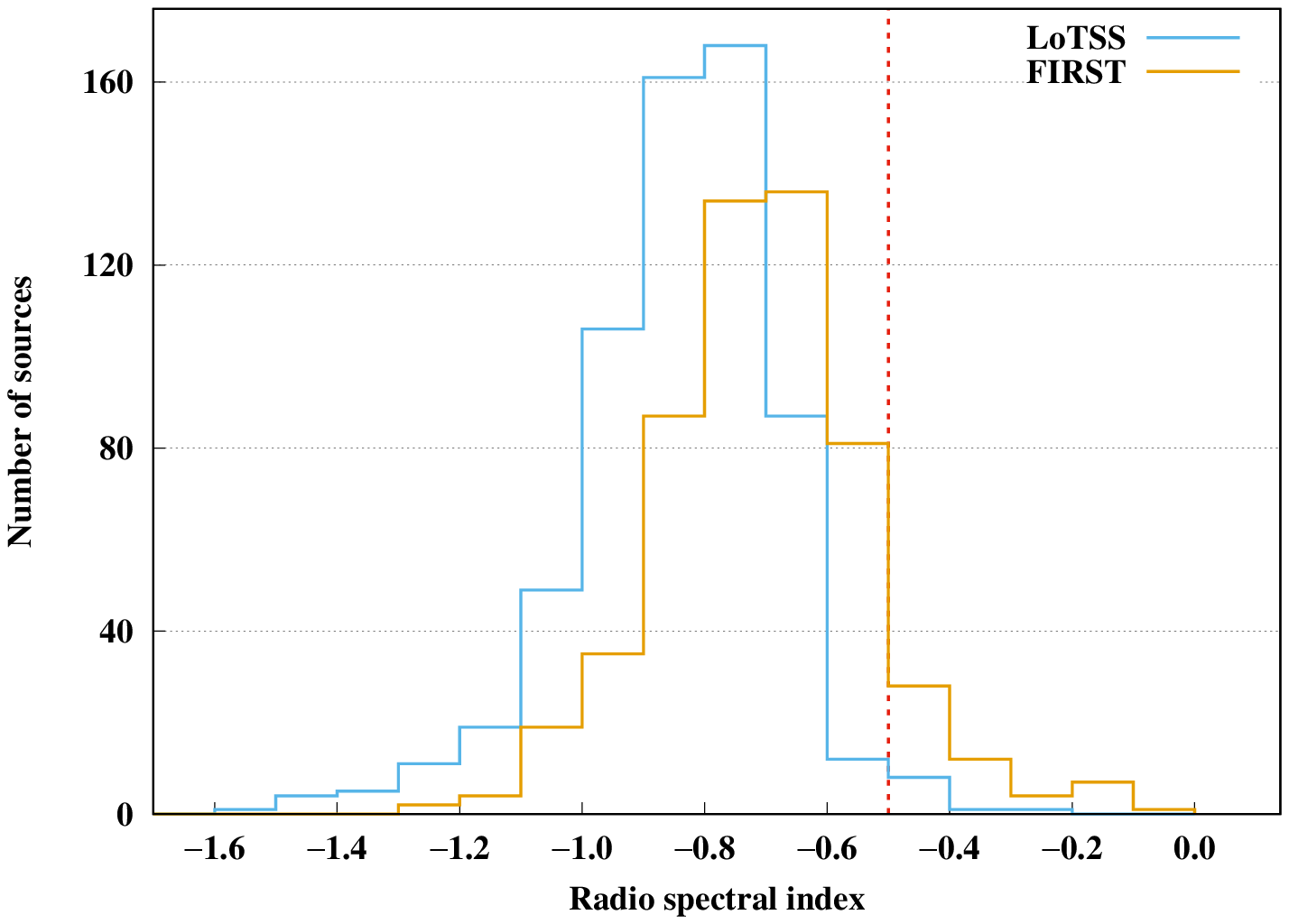}
\caption{Spectral index distribution of the newly identified winged radio sources (left panel) and the FIRST and LoTSS samples (right panel). The vertical red dashed line ($\alpha$ = --0.50) marks the boundary between steep spectra (left of the line) and flat spectra (right of the line). The bin size for both plots is set to 0.1.}
\label{fig:spindex_histgrm}
\end{figure*}

\subsection{Spectral Index}
We estimated the two-point spectral index for sources from both the FIRST and LoTSS samples using LoTSS (144 MHz) and NVSS (1400 MHz) flux measurements for the LoTSS samples and TGSS (150 MHz) and NVSS (1400 MHz) flux measurements for the FIRST samples. The NVSS flux was chosen instead of FIRST due to differences in observation configurations. The NVSS observations, performed in the VLA D configuration, provide more accurate flux measurements, even though they may slightly compromise positional accuracy. In contrast, the high resolution and lack of short-spacing antennas in the FIRST B configuration can result in missing flux for extended sources.

We focused first on the winged sources presented in this paper. The spectral index was calculated for 607 of the 621 sources, as NVSS flux measurements were unavailable for the remaining 14 cases. A histogram showing the spectral index distribution is presented in Figure \ref{fig:spindex_histgrm} (left panel). Nearly all sources exhibit steep spectra ($\alpha$ $\leq$ --0.50), with only 9 sources ($\sim$1.5\%) displaying flat spectra. The predominance of steep spectra is consistent with expectations for lobe-dominated radio sources and the overall spectral index value is --0.84$\pm$0.16. Among the 607 data points, XRGs account for 375, while ZRGs contribute 232. The mean spectral index values are --0.86$\pm$0.15 and --0.80$\pm$0.18 for XRGs and ZRGs, respectively. This observed trend of XRGs having steeper spectra than ZRGs aligns with previous findings \citep{Be20, Be22}. It may be attributed to the presence of more diffuse wings in XRGs compared to ZRGs, which influences their spectral characteristics.

Next, we compared the spectral index distribution of winged sources from the FIRST and LoTSS samples (Figure \ref{fig:spindex_histgrm}, right panel). Since many sources in the FIRST sample lack LoTSS measurements at 144 MHz, we used their measurements from the TGSS catalog at 150 MHz \citep{In17} or derived them directly from TGSS observations. Consequently, the histogram in Figure \ref{fig:spindex_histgrm} represents $\alpha_{150}^{1400}$ for FIRST sources and $\alpha_{144}^{1400}$ for LoTSS sources. We assume that $\alpha_{150}^{1400}$ and $\alpha_{144}^{1400}$ depict broadly similar spectral characteristics. The dataset comprises 550 spectral index measurements from FIRST and 633 from LoTSS. Among all winged sources, the majority ($\sim$65\%) exhibit spectral indices within the range of --0.60 $\leq \alpha \leq$ --0.90. Additionally, most winged sources ($\sim$95\%) have steep spectra, consistent with the expectation for lobe-dominated radio sources. From Figure \ref{fig:spindex_histgrm}, we observe a slight difference in the mean spectral index between the FIRST and LoTSS samples, with values of --0.70 and --0.84, respectively. This discrepancy might stem from the advantage of identifying winged sources at lower frequencies, where diffuse emission becomes more prominent.

Radio galaxies typically exhibit spectral indices in the range of --0.7 to --0.8, a characteristic associated with their synchrotron emission \citep{Gi82, Kl18}. The average spectral index values for WRGs are consistent with those of typical radio galaxies, reaffirming their shared physical nature.

\begin{figure*}
\includegraphics[angle=-90,width=9.0cm]{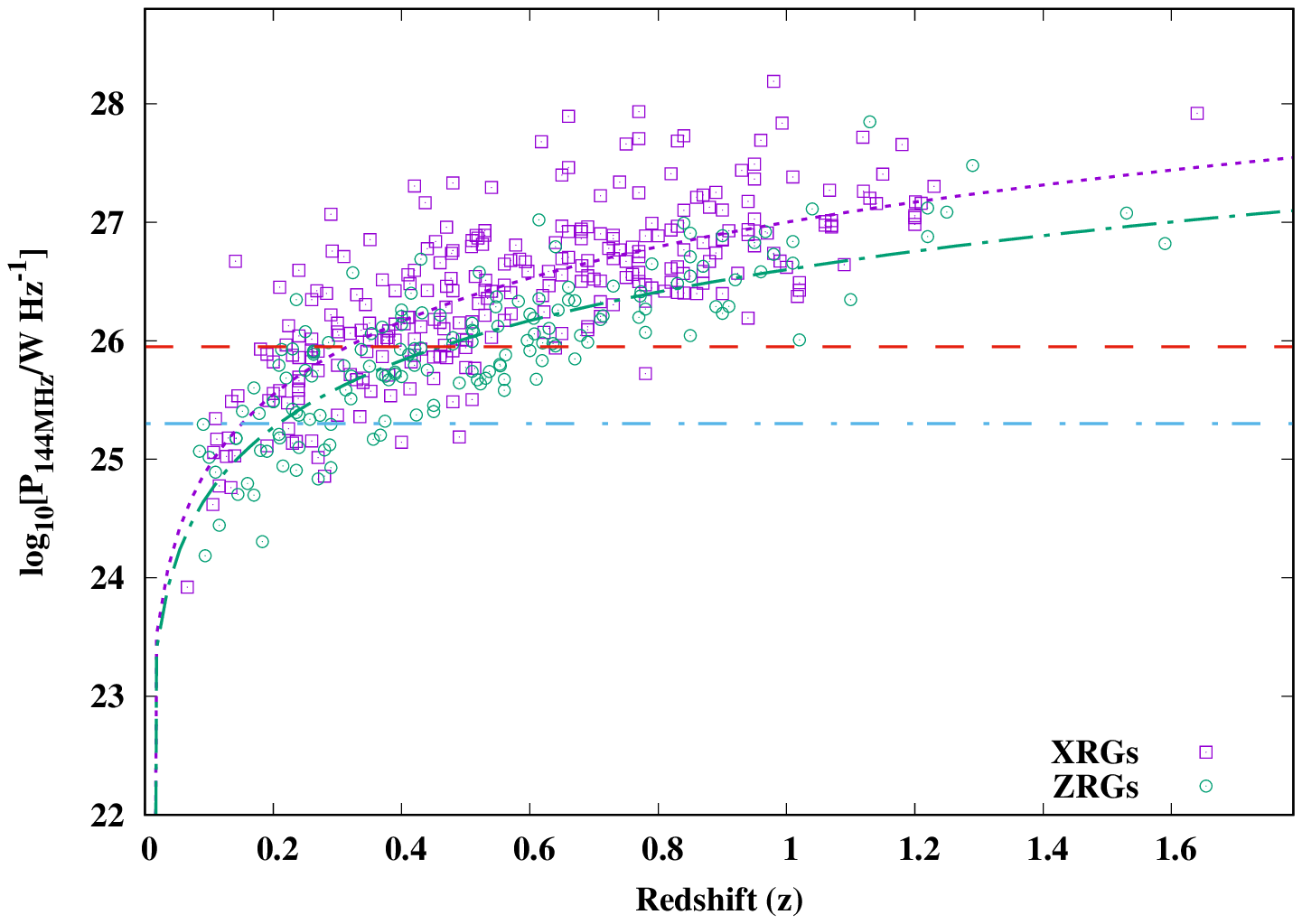}
\includegraphics[angle=-90,width=9.0cm]{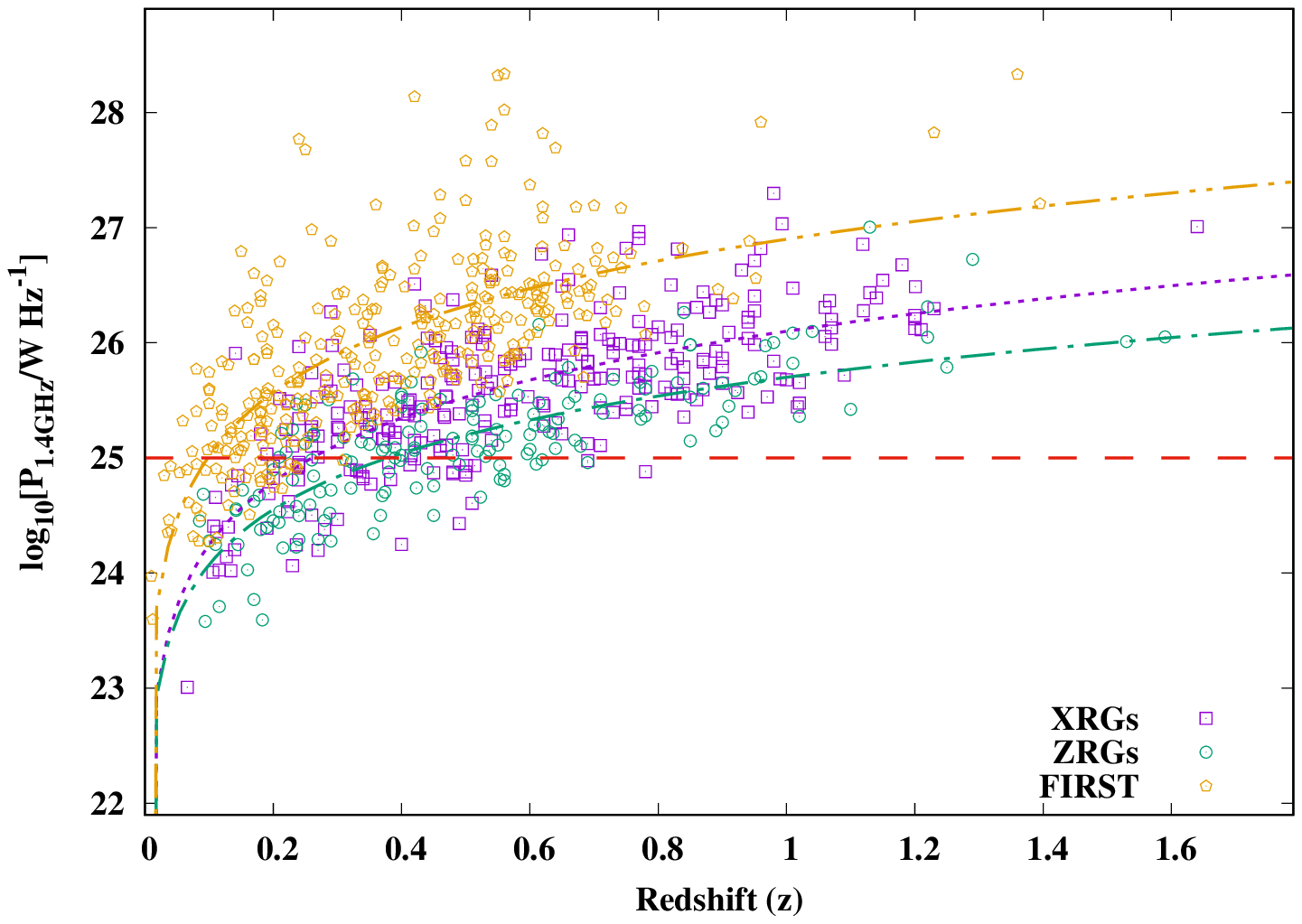}
\caption{Distribution of radio power versus redshift at 144 MHz (left panel) and 1.4 GHz (right panel) for the newly identified winged sources and the FIRST sample. The curved lines represent the fitted power trends for XRGs (violet line), ZRGs (green line), and the FIRST sample (orange line). Horizontal lines indicate the median or mean power values (as described in the text) for FR-I sources (blue dashed-dot line) and FR-II sources (red dashed line).}
\label{fig:redshiftpower}
\end{figure*}

\subsection{Radio Power}
The radio power is estimated using the following formula, adopted from \citet{Do09} :
\begin{equation}
P_{\nu} = 4{\pi}D^2_{L}F_{\nu}(1 + z)^{-(1+{\alpha})}
\end{equation}
when the symbols are: $\rm D_{L}$ -- the luminosity distance, $\rm F_{\nu}$ -- the radio ﬂux at a frequency $\nu$, $\rm (1 + z)^{-(1+{\alpha})}$ is the standard k-correction used in radio astronomy \citep{Ho02}. The radio power at 144 MHz ($\rm P_{144MHz}$) was calculated for 441 winged sources (281 XRGs and 160 ZRGs) from this paper, using available spectral index $\alpha_{144}^{1400}$ values and redshift information. The average radio power for the entire group at 144 MHz is $\rm log_{10}[P_{144MHz}/W Hz^{-1}]$=26.25, while for individual classes--XRGs and ZRGs--it is 26.43 and 25.93, respectively.

In Figure \ref{fig:redshiftpower} (left panel), we present the distribution in radio power with redshift for XRGs and ZRGs. While both groups span a similar redshift range, their radio powers differ significantly, with ZRGs generally being less powerful than XRGs. This discrepancy likely arises from the lower flux and spectral values of ZRGs compared to XRGs, although the exact reason remains unclear. Additionally, we have included two horizontal lines indicating the median 144 MHz radio power of FR-I and FR-II sources as reported by \citet{Mi19} based on LoTSS DR1 observations. These lines show that the majority of winged sources have luminosities close or well above the FR-II median value, classifying them predominantly as FR-II sources.

In Figure \ref{fig:redshiftpower} (right panel), we plot the distribution of radio power at 1.4 GHz versus redshift for XRGs, ZRGs, and objects from the FIRST sample. As previously observed, XRGs are more luminous than ZRGs. Objects from the FIRST sample exhibit even higher luminosities on average, which can be attributed to the observational bias already mentioned here. The mean radio power at 1.4 GHz for the FIRST sample is $\rm log_{10}[P_{1.4GHz}/W Hz^{-1}]$=25.94, compared to 25.59 for XRGs and 25.14 for ZRGs.

\begin{figure}
\includegraphics[angle=-90,width=9.0cm]{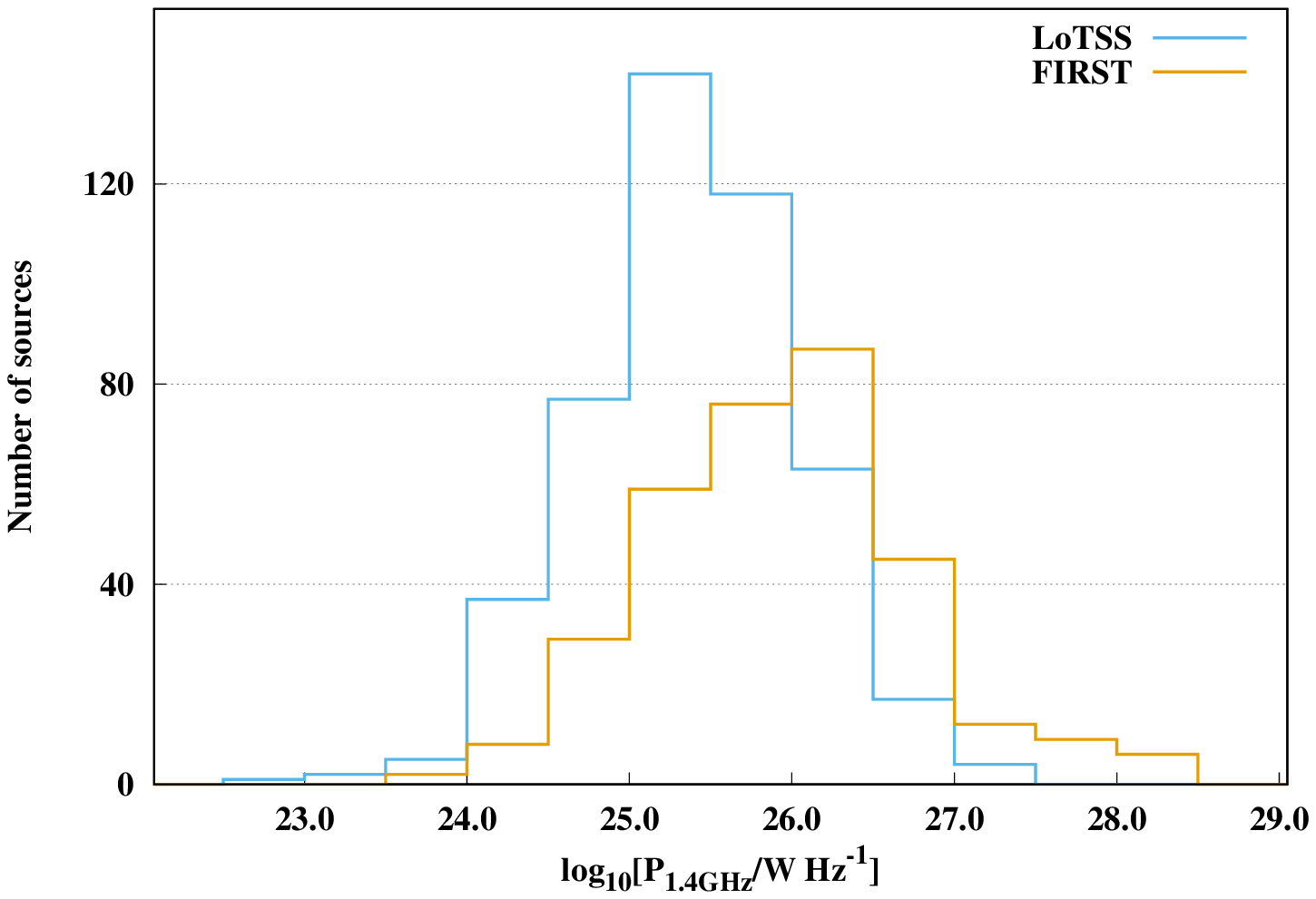}
\caption{Radio power distribution at 1.4 GHz for winged sources from FIRST and LoTSS samples. The bin size is set to 0.5.}
\label{fig:powerAll}
\end{figure}

This observation is further supported by the histogram in Figure \ref{fig:powerAll}, which depicts the distribution of $\rm P_{1.4GHz}$ for the LoTSS and FIRST samples. The average radio power for the LoTSS sample is $\rm log_{10}[P_{1.4GHz}/W Hz^{-1}]$=25.39. As indicated in Figure \ref{fig:redshiftpower} (right panel), FR-II sources typically have $\rm log_{10}[P_{1.4GHz}/W Hz^{-1}] \ge$25 \citep{Ow94, Ko11}. Thus, the radio power of winged sources is statistically indistinguishable from that of regular FR-II radio galaxies. This suggests that the underlying AGN mechanisms powering the primary jets may not differ significantly between these two classes, despite the presence of additional features, such as the wings. However, further detailed studies are required to confirm this hypothesis.

For the giant winged sources, the mean radio power at 1.4 GHz is $\rm log_{10}[P_{1.4GHz}/W Hz^{-1}]$=25.56. This value is comparable to that for giant radio sources ($\rm log_{10}[P_{1.4GHz}/W Hz^{-1}]$=25.5; \citet{Ku18}). A detailed investigation is necessary to determine whether a similar intrinsic radio emission mechanism at the core underpins both classes of sources.

\begin{figure*}
\includegraphics[angle=-90,width=9.0cm]{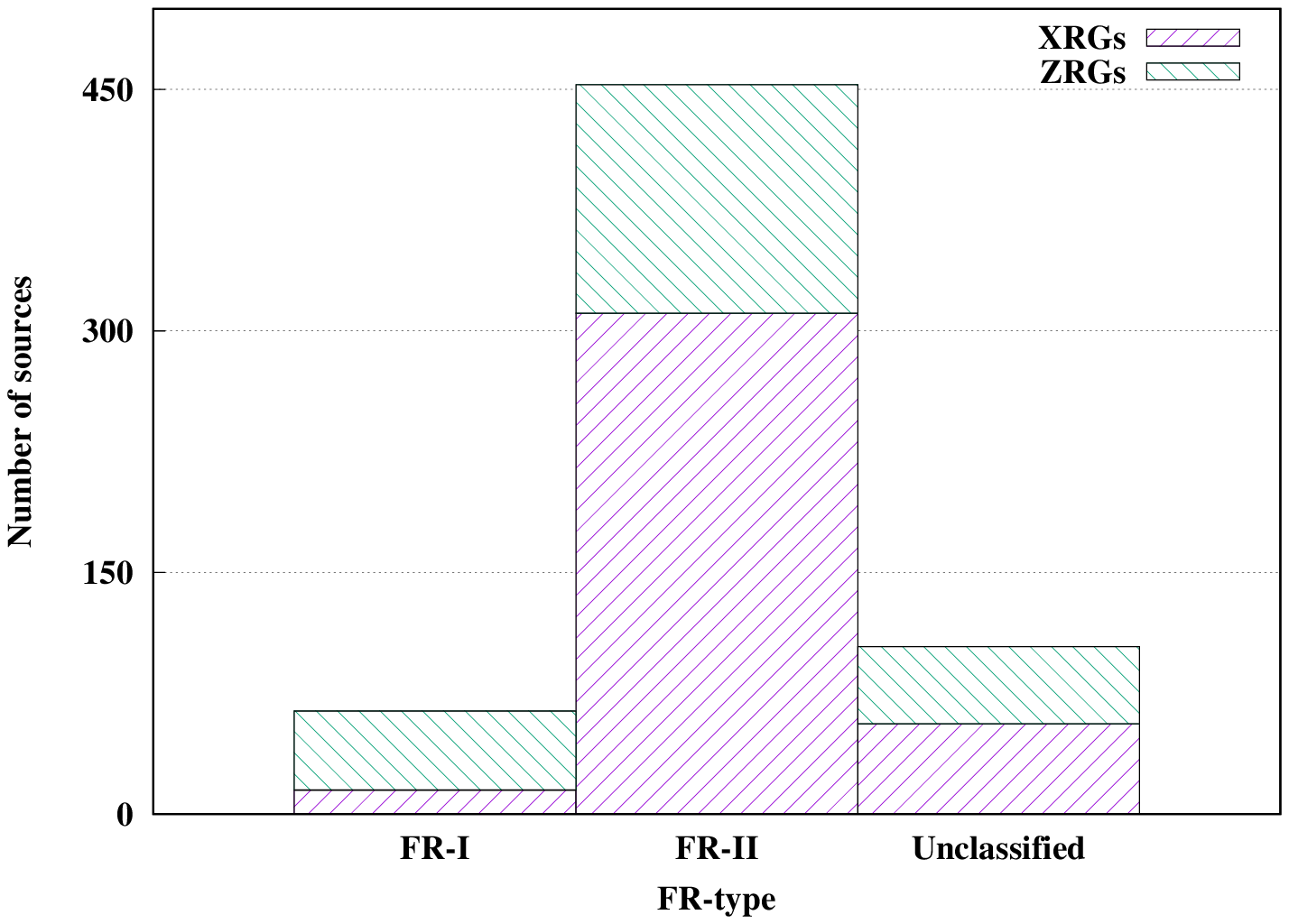}
\includegraphics[angle=-90,width=9.0cm]{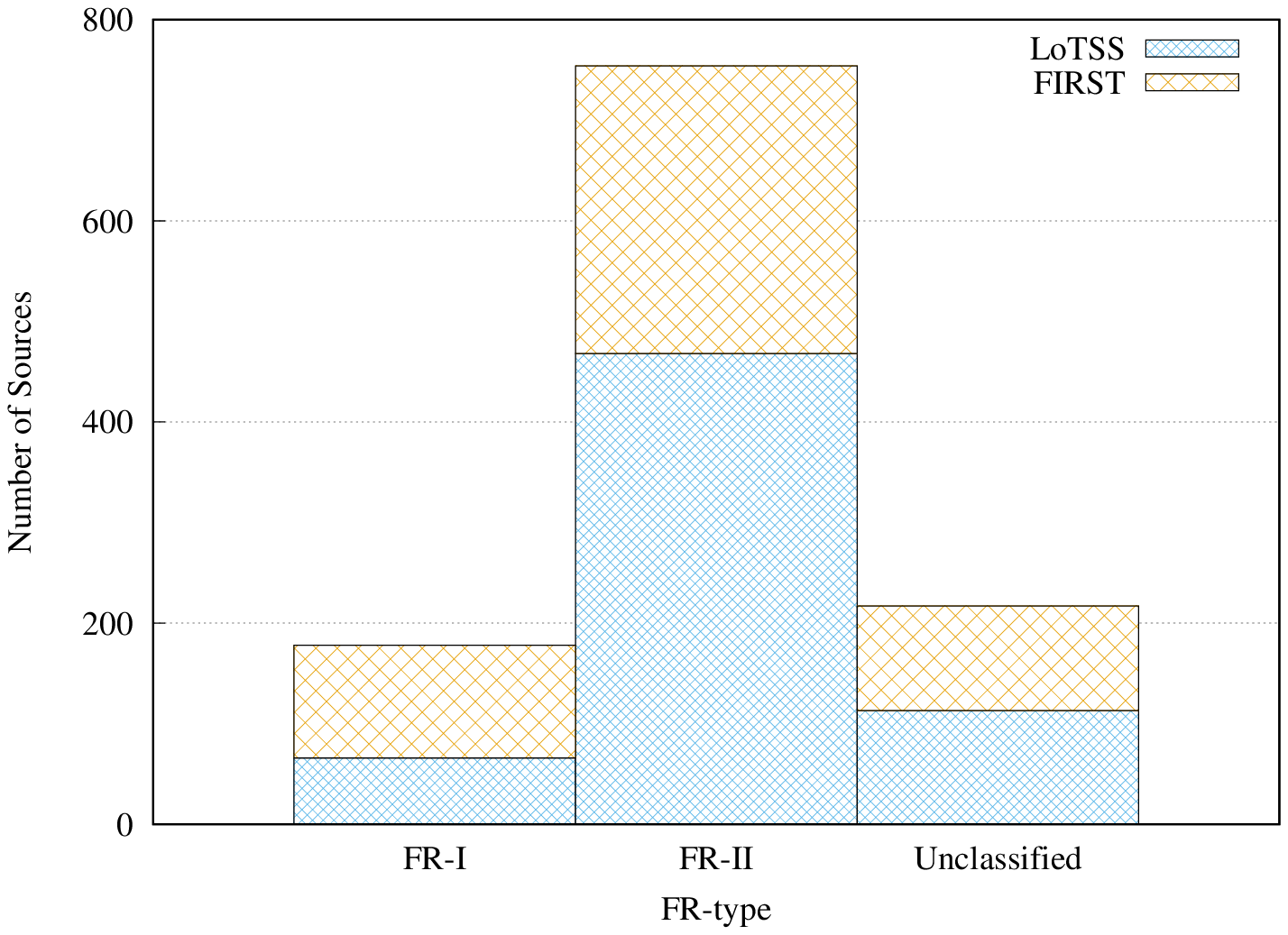}
\caption{The FR-classification of newly identified winged sources, divided into XRGs and ZRGs (left panel), along with the FR-classification of winged sources from the FIRST and LoTSS samples (right panel).}
\label{fig:frtype}
\end{figure*}

\subsection{FR dichotomy}
We have examined the Fanaroff-Riley (FR) classification \citep{Fa74} for the newly identified winged sources. As previously noted, the FR-classification is based purely on radio morphology and was introduced over 50 years ago. Since its inception, this system has been widely applied to various catalogs and populations of radio sources. Over time, our understanding of the relationship between FR-classification and the dynamics of radio sources, as well as the AGN fueling mechanisms, has significantly advanced. There is evidence suggesting a connection between accretion mode and radio morphology \citep{Ge13, Mi22}. However, the exact physical conditions necessary to produce the two FR classes remain unclear. Edge-brightened FR-II sources are believed to host jets that stay relativistic throughout, terminating in hotspots (internal shocks), while edge-dimmed FR-I sources are thought to have jets that disrupt on kiloparsec scales.

To classify the winged sources, we assessed the relative brightness between the core and edges and identified the position of the peak brightness along the major axis. Based on this analysis, 64 sources were classified as FR-I, and 453 were identified as FR-II. Among the FR-I sources, 15 are XRGs and 49 are ZRGs. Similarly, the FR-II category includes 311 XRGs and 142 ZRGs. A total of 104 sources could not be classified due to ambiguous brightness distributions, which may indicate these sources are candidates for HyMORS. The left panel of Figure \ref{fig:frtype} shows the FR classifications as a histogram. Among the classified sources, the majority ($\sim$88\%) are FR-II, consistent with their high radio power. Notably, $\sim$95\% of classified XRGs are FR-II, supporting previous findings that X-shaped sources are predominantly found in FR-II galaxies \citep{Le92, De02}. This aligns with results from \citet{Sa18}, where only 2.7\% (1 of 37) of XRGs were classified as FR-I. Interestingly, the FR-I category is dominated by ZRGs, whereas the number of FR-I XRGs is negligible. This suggests that edge-darkened winged sources are more likely to exhibit a Z-shape rather than an X-shape.

The right panel of Figure \ref{fig:frtype} presents the distribution of winged sources with their respective FR-classifications for both the FIRST and LoTSS samples. For this comparison, we classified previously identified winged sources primarily from \citet{Ch07}, \citet{Ya19}, and \citet{Be20, Be22}. Out of 1,149 winged sources, we identified 178 as FR-I, 754 as FR-II, and 217 as unclassified. Among the classified sources, the majority ($\sim$81\%) are FR-II, consistent with our findings for the LoTSS DR2 sample. Additionally, most classified XRGs belong to the FR-II category, while the FR-I category is primarily composed of ZRGs. In Figure \ref{fig:sampleoffr} (provided in the Appendix), we present examples of sources classified as FR-I and FR-II, as well as those for which unequivocal classification was not possible.

\section{Summary}
\label{sec:summary}
Given the critical importance of morphological studies for understanding radio galaxies, we are undertaking a large-scale effort to identify and analyze various morphological classes and subclasses using low-frequency and high-resolution data. This paper represents the first installment of our findings, focusing on the identification of new winged radio sources from the LoTSS DR2 survey.

In total, we present 1,024 newly identified winged radio sources. This includes 621 confirmed winged radio sources and 403 candidates. Among the confirmed sources, 382 are classified as X-shaped radio galaxies (XRGs), and 239 are classified as Z-shaped radio galaxies (ZRGs).

While the primary focus of this paper is the identification of winged radio sources, we also provide an initial analysis of their parameters and properties. The general characteristics of the winged sources can be summarized as follows:

$\bullet$ The median angular and linear sizes of winged sources are 82 arcseconds and 498 kpc, respectively. However, approximately 16\% of them have linear sizes exceeding 0.7 Mpc, classifying them as potential giant radio galaxies.

$\bullet$ The winged sources predominantly exhibit steep spectra between 1.4 GHz and 144 MHz, with a mean spectral index of --0.84. However, when divided into XRG and ZRG categories, we observe that XRGs have steeper spectra than ZRGs. This may be due to the presence of more diffuse wings in XRGs compared to ZRGs, which influences their spectral characteristics. When we further compare the spectra of winged sources identified in this study with those identified previously based on FIRST 1.4 GHz observations, we find that the former have slightly steeper spectra. The mean spectral index for winged sources identified using FIRST 1.4 GHz data is --0.70. This difference can be attributed to the superior sensitivity and resolution of the LoTSS DR2 data, which enables the identification of fainter sources.

$\bullet$ The distribution of radio power for winged sources at both studied frequencies, 1.4 GHz and 144 MHz, shows that the majority of them have powers indistinguishable from those of regular FR-II radio galaxies. Additionally, XRGs are found to be more luminous compared to ZRGs.

$\bullet$ Based on the FR classification, we conclude that the majority of winged sources ($\sim$88\%) exhibit the edge-brightened morphology typical of FR-II sources, and this percentage is even higher ($\sim$95\%) for the XRG class alone. This is consistent with their radio power values observed.

A comprehensive study will be presented in forthcoming works, focusing on the optical and radio properties of these winged sources. Additionally, results for other classes and subclasses of radio sources will also be detailed in subsequent publications.

\section*{Acknowledgments}
We are thankful to the anonymous referee for the  constructive comments that helped to improve this paper. This work is supported by the National Natural Science Foundation of China under Nos. 11890692, 12133008, 12221003. S.K.B. and T.F. acknowledge the science research grant from the China Manned Space Project with No. CMS-CSST-2021-A04. T.K.S. and X.C. acknowledge funding support from the National SKA Program of China (No. 2022SKA0110100 and 2022SKA0110101). MKB acknowledges support from the `National Science Centre, Poland' under grant no. 2017/26/E/ST9/00216.

\section{Appendix}

\begin{table*}
\caption{The candidates of winged radio sources from LoTSS DR2}
\label{table:candidate}
\begin{centering}
\scriptsize
\begin{tabular}{ccllccccccccl}
\hline\hline
Cat & Short      &~~~~~R.A.      &~~~~~Decl.   & $z$    &$z_{type}$&$m_{r}$&$M_{R}$&$F_{144}$&Isl\_rms&$\theta$&$l$  &Comment(s)\\
No. & Name       &~~(J2000.0)    &~~(J2000.0)  &        &          &       &       &  (mJy)  &(mJy/   &(arcsec)&(kpc)&          \\
    &            &               &             &        &          &       &       &         &beam)   &        &     &          \\
\hline
~~1 & J0001+3607 & 00 01 39.0   & +36 07 31$^{\ast}$  &  ---   &   ---   &  ---  &  ---   &~329.0  & 0.187 & ~93   &  ---  & one sided and small wing                      \\
~~2 & J0004+1946 & 00 04 04.3   & +19 46 50$^{\ast}$  &  ---   &   ---   &  ---  &  ---   &~190.3  & 0.121 & ~99   &  ---  & one sided and small wing                      \\
~~3 & J0004+4006 & 00 04 19.4   & +40 06 12$^{\ast}$  &  ---   &   ---   &  ---  &  ---   &~372.3  & 0.097 & ~60   &  ---  & one sided and small wing                      \\
~~4 & J0005+3751 & 00 05 44.0   & +37 51 49$^{\ast}$  &  ---   &   ---   &  ---  &  ---   &~~45.1  & 0.098 & ~51   &  ---  & significant on one side and another side small wing \\ 
~~5 & J0010+3045 & 00 10 06.2   & +30 45 26$^{\ast}$  &  ---   &   ---   &  ---  &  ---   & 2354.2 & 0.183 & ~79   &  ---  & one sided and significant wing                \\
\hline
\end{tabular}
\end{centering}
\scriptsize
Notes: This is a sample excerpt from the table, which will be available in its entirety in the online link of the published version.\\
Column information of the Table: The Column 1 is the catalog number. The short name (IAU source identifier as JHHMM+DDMM) of the candidate sources is presented in Column 2. Columns 3 and 4 are the R.A. (in hh mm ss.ss) and Decl. (in dd mm ss.s), both in J2000.0. Like the previous table, primarily we have used the optical position, and radio positions are given if the optical information is unavailable. The radio positions are given with less precision (R.A. as hh mm ss.s and Decl. as dd mm ss) The redshift ($z$) and redshift ($z_{type}$) types are presented in Columns 5 and 6, respectively. Similar to the previous table, the spectroscopic redshift values are provided either from the Sloan Digital Sky Survey Data Release 16 (SDSS DR16: \citet{Ah20}) or the Dark Energy Spectroscopic Instrument (DESI: \citet{Le13}), when the photometric values are obtained from \citet{Du22}. The PHOT redshifts are presented with less precision, up to 0.01 place. In the next two columns, we have listed r-band magnitude ($m_{r}$) and r-band rest frame magnitude ($M_{R}$). The LoTSS DR2 flux ($F_{144}$ in mJy) of the sources is given in Column 9. Column 10 lists each source's corresponding rms noise in the island (Isl\_rms). The angular size ($\theta$) and the linear size ($l$) of the sources are presented in Column 11 and Column 12, respectively. In the last column, we put comments about the source, which mainly indicates the nature of their wings, e.g., if the sources have a one-sided wing, apparently small size wing, no prominent wing morphology, etc.\\
$^{\dag}$ The source is present in \citet{Pr11}.\\
The coordinates with $^{\ast}$ marks indicate that they are radio coordinates; otherwise, they are the respective optical host position.\\
The $^{\star}$ mark on the spectroscopic redshift implies that the redshift is taken from DESI; otherwise, it is taken from SDSS.
\end{table*}
 \begin{table*}
\caption{List of previously identified winged sources also detected in LoTSS DR2}
\label{table:common}
\begin{centering}
\hspace*{0.01cm}
\begin{tabular}{cccl||cccl}
\hline\hline
 Short      & R.A.         & Decl.        & References  & Short      & R.A.         & Decl.        & References\\
 Name       &(J2000.0)     &(J2000.0)     &             & Name       &(J2000.0)     &(J2000.0)     &           \\
\hline                                                    
 J0058+2651 &  00 58 22.63 &  +26 51 58.7 &  Ek78, Ch07 & J1302+5119 &  13 02 58.46 &  +51 19 43.6 &  Ya19, Be22\\
 J0724+3803 &  00 58 22.64 &  +26 51 58.6 &  Bh22	& J1310+4644 &  13 10 42.38 &  +46 44 35.0 &  Ya19, Be22\\
 J0741+3333 &  07 41 25.22 &  +33 33 19.9 &  Ya19	& J1310+5458 &  13 10 15.40 &  +54 58 34.2 &  Ch07, Be22\\
 J0738+3846 &  07 38 54.81 &  +38 46 27.8 &  Be20	& J1314+5439 &  13 14 04.60 &  +54 39 37.9 &  Be22     	\\
 J0823+5812 &  08 23 33.47 &  +58 12 11.0 &  Ya19	& J1315+5254 &  13 15 31.08 &  +52 54 37.6 &  Be22     	\\
 J0831+3219 &  08 31 27.49 &  +32 19 26.8 &  Pa85, Ch07 & J1316+2427 &  13 16 38.27 &  +24 27 32.4 &  Ch07     	\\
 J0834+6635 &  08 34 09.01 &  +66 35 48.9 &  Bh22	& J1324+5041 &  13 24 35.20 &  +50 41 02.3 &  Be22     	\\
 J0836+3125 &  08 36 35.46 &  +31 25 51.2 &  Ch07	& J1336+4900 &  13 36 14.99 &  +49 00 04.8 &  Be22     	\\
 J0837+4450 &  08 37 52.75 &  +44 50 25.8 &  Ya19	& J1336+3626 &  13 36 27.73 &  +36 26 27.3 &  Be20     	\\
 J0838+3253 &  08 38 44.61 &  +32 53 11.8 &  Ch07	& J1340+5035 &  13 40 02.96 &  +50 35 39.7 &  Ya19, Be22\\
 J0845+4031 &  08 45 08.40 &  +40 31 15.4 &  Ch07	& J1341+2622 &  13 41 53.15 &  +26 22 48.8 &  Bh22     	\\
 J0929+3121 &  09 29 54.13 &  +31 21 28.3 &  Ya19	& J1345+5233 &  13 45 41.64 &  +52 33 35.6 &  Ch07, Be22\\
 J0941+3944 &  09 41 24.03 &  +39 44 41.8 &  Bl92, Ch07 & J1345+5403 &  13 45 57.55 &  +54 03 16.6 &  Be22     	\\
 J0943+2834 &  09 43 02.25 &  +28 34 45.8 &  Ch07	& J1346+5410 &  13 46 32.69 &  +54 10 31.6 &  Be22     	\\
 J0949+4456 &  09 49 53.64 &  +44 56 55.7 &  Ya19	& J1351+5559 &  13 51 42.14 &  +55 59 43.1 &  Ch07, Be22\\
 J1020+4831 &  10 20 51.92 &  +48 31 09.8 &  Va82, Ch07 & J1357+4807 &  13 57 30.5  &  +48 07 41   &  Le01, Ch07\\
 J1021+4425 &  10 21 16.98 &  +44 25 39.8 &  Be20	& J1359+4601 &  13 59 08.72 &  +46 01 13.7 &  Be22     	\\
 J1039+4648 &  10 39 24.92 &  +46 48 11.5 &  Ya19	& J1403+4953 &  14 03 49.79 &  +49 53 05.4 &  Ya19, Be22\\
 J1049+4422 &  10 49 35.28 &  +44 22 04.0 &  Ch07	& J1416+5425 &  14 16 29.03 &  +54 25 32.1 &  Be22     	\\
 J1054+5521 &  10 54 00.60 &  +55 21 53.0 &  Ch07, Be22 & J1430+5217 &  14 30 17.34 &  +52 17 35.0 &  Ch07, Be22\\
 J1054+4703 &  10 54 26.39 &  +47 03 27.4 &  Ya19, Be22 & J1442+5043 &  14 42 19.18 &  +50 43 57.9 &  Be22     	\\
 J1056+5112 &  10 56 44.29 &  +51 12 14.2 &  Be22	& J1444+4147 &  14 44 07.25 &  +41 47 50.3 &  Ch07     	\\
 J1112+4755 &  11 12 17.58 &  +47 55 56.6 &  Be22	& J1446+4831 &  14 46 29.46 &  +48 31 54.1 &  Be22     	\\
 J1115+5600 &  11 15 29.59 &  +56 00 39.6 &  Be22	& J1453+5317 &  14 53 21.20 &  +53 17 52.3 &  Be22     	\\
 J1121+5344 &  11 21 26.44 &  +53 44 56.7 &  Be22	& J1502+5304 &  15 02 09.53 &  +53 04 19.9 &  Be22     	\\
 J1129+5407 &  11 29 19.35 &  +54 07 33.9 &  Be22	& J1502+5244 &  15 02 29.04 &  +52 44 02.1 &  Be22     	\\
 J1132+5558 &  11 32 22.74 &  +55 58 18.5 &  Be22	& J1504+5749 &  15 04 08.06 &  +57 49 22.6 &  Be20     	\\
 J1138+4950 &  11 38 16.62 &  +49 50 25.0 &  Ya19, Be22 & J1508+6137 &  15 08 16.29 &  +61 37 56.3 &  Ya19     	\\
 J1139+5312 &  11 39 56.83 &  +53 12 11.8 &  Ya19, Be22 & J1519+5007 &  15 19 33.73 &  +50 07 25.0 &  Be22     	\\
 J1142+5800 &  11 42 15.26 &  +58 00 01.9 &  Ya19	& J1519+5342 &  15 19 36.72 &  +53 42 55.4 &  Be20, Be22\\
 J1149+4618 &  11 49 50.67 &  +46 18 50.6 &  Be20, Be22 & J1530+3301 &  15 30 22.23 &  +33 01 19.5 &  Be20     	\\
 J1154+4835 &  11 54 18.72 &  +48 35 21.1 &  Be22	& J1540+4602 &  15 40 09.75 &  +46 02 20.1 &  Ya19     	\\
 J1155+4417 &  11 55 00.34 &  +44 17 02.2 &  Ya19	& J1544+3044 &  15 44 13.39 &  +30 44 01.1 &  Ya19     	\\
 J1202+4915 &  12 02 35.10 &  +49 15 31.7 &  Ch07, Be22 & J1548+4451 &  15 48 17.19 &  +44 51 47.4 &  Ya19     	\\
 J1216+5243 &  12 16 23.68 &  +52 43 59.9 &  Be22	& J1617+3222 &  16 17 42.53 &  +32 22 34.3 &  Be20     	\\
 J1224+5623 &  12 24 37.88 &  +56 23 40.0 &  Be22	& J1639+5546 &  16 39 50.31 &  +55 46 10.6 &  Bh22     	\\
 J1239+5314 &  12 39 15.39 &  +53 14 14.6 &  Be22	& J1709+3425 &  17 09 39.15 &  +34 25 50.8 &  Be20     	\\
 J1243+5212 &  12 43 08.97 &  +52 12 45.0 &  Be22	& J2307+1920 &  23 07 26.50 &  +19 20 44.8 &  Bh22     	\\
 J1258+3227 &  12 58 32.87 &  +32 27 40.8 &  Ch07	& J2336+2108 &  23 36 30.48 &  +21 08 46.9 &  Bh22     	\\ 
\hline        
\end{tabular} 
\end{centering}
\\
\vskip 0.01cm
\begin{centering}
\hskip 0.60cm {References-- Be20: \citet{Be20}; Be22: \citet{Be22}; Bh22: \citet{Bh22}; Bl92: \citet{Bl92}}\\
\hskip 2.15cm Ch07: \citet{Ch07}; Ek78: \citet{Ek78}; Le01: \citet{Le01}; Pa85: \citet{Pa85}; \\
\hskip 3.56cm Va82: \citet{Va82}; Ya19: \citet{Ya19}
\end{centering}
\end{table*}
 
\begin{figure*}
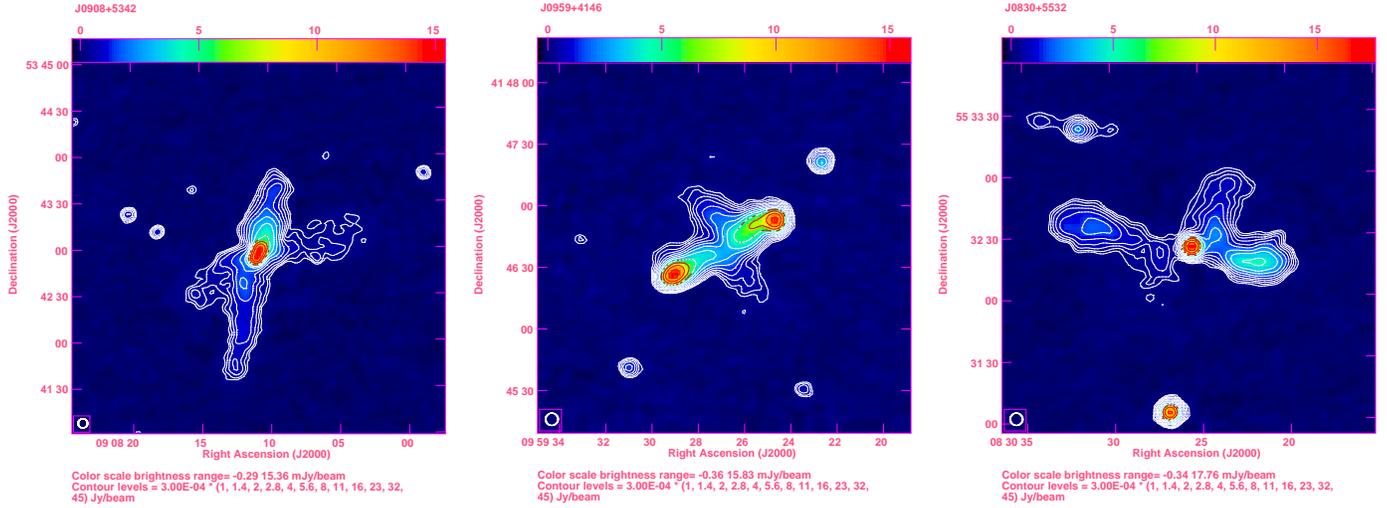

\vbox{
\centerline{
\includegraphics[angle=0,height=7.0cm,width=6.1cm]{./J0908+5342.PS}
\includegraphics[angle=0,height=7.0cm,width=6.1cm]{./J0959+4146.PS}
\includegraphics[angle=0,height=7.0cm,width=6.1cm]{./J0830+5532.PS}}}
\caption{Examples of radio images of sources classified as (from left to right): FR-I (J0908+5342), FR-II (J0959+4146) and those for which no classiﬁcation was made (J0830+5532).}
\label{fig:sampleoffr}
\end{figure*}

\clearpage

\bibliographystyle{aasjournal}

\end{document}